\documentclass[modern]{aastex62}
\usepackage{amsmath}
\usepackage{ulem}
\usepackage{CJK}

\accepted{in AJ, 2019 October 8}

\shorttitle{Dependence of Lightcurves on Phase \& Shape}
\shortauthors{Lu \& Jewitt}

\begin{document}
\title{Dependence of Lightcurves on Phase Angle and Asteroid Shape}

\correspondingauthor{Xiao-Ping Lu}
\email{xplu@must.edu.mo}

\author[0000-0002-2363-4175]{Xiao-Ping Lu}
\affil{Faculty of Information Technology, Macau University of Science and Technology, Taipa, Macao, China}
\affil{State Key Laboratory of Lunar and Planetary Sciences(Macau University of Science and Technology)}

\author{David Jewitt}
\affiliation{Department of Earth, Planetary and Space Sciences, UCLA, 595 Charles Young Drive East, Los Angeles, CA 90095-1567, USA}
\affiliation{Department of Physics and Astronomy, UCLA, 430 Portola Plaza, Box 951547, Los Angeles, CA 90095-1547, USA}

\begin{abstract}

We investigate the phase angle dependence of asteroid lightcurves  using numerical scattering models applied to simple body shapes.  For simplicity, the Kaasalainen scattering parameters are  obtained from the corresponding Hapke scattering laws for C-type, S-type, and V-type asteroids. The spectral types differ substantially in the role of multiple scattering (which is largely a function of their  geometric albedos) but we find that the differences on the light curve vs.~phase relations are modest. By using a Kaasalainen scattering law, the amplitudes and axis ratios with respect to different  phase angles from $0^\circ$ to $140^\circ$ are plotted for these types asteroids based on a  biaxial ellipsoid shape model. Additionally, we examine the relationship between amplitude and axis ratio for a contact binary  represented by identical biaxial ellipsoids, including the effects of shadowing of one component by the other. We compare the models with published high phase angle observations, and with interstellar object  1I/`Oumuamua ($\Delta m = 2.5$ magnitude at phase angle $\alpha = 23^\circ$), finding axis ratio $5.2:1$ if represented as a single ellipsoid and  $3.5:1$ for each component if represented as  a nose-to-nose  contact binary. While a detailed fit is not attempted, the  comparison shows that the single ellipsoid model is better.
\end{abstract}



\section{Introduction} \label{sec:intro}

A huge number of photometric observations of asteroids  are obtained everyday, greatly enriching our knowledge of this population. The most basic parameters to be extracted from photometric data include estimates of the asteroid size (if the albedo is known or can be assumed) and the shape.  Of course, while the true shapes of asteroids are irregular, it is convenient to represent them by geometric forms, typically as triaxial ellipsoids characterized by their axis ratios.  The axis ratios can then be inferred from observed brightness variations as the body rotates.  \\

However, in addition to the intrinsic shape, it is well known that the light curve shape and amplitude also depend on the angle between the spin vector and the line of sight, and  on the phase (Sun-object-observer) angle, $\alpha$. Generally the amplitude grows larger (because of scattering and self-shadowing effects), for a given axis ratio, as the phase angle increases.  \cite{Zappala1990} first analyzed the relationship between light curve  amplitude and solar phase angle among asteroids. They found that the amplitude is linearly proportional to phase angle for  $\alpha \le 50^\circ$. The phase angle dependence of the light curve has also been investigated by \cite{2007AJ_Jewitt} but has otherwise not received much quantitative attention. \\

In this paper we investigate the relationship between amplitude,  axis ratio, spectral type and phase angle based on  biaxial ellipsoid models, in order to provide a quick and practical  tool for observers.  The numerical method of calculating the amplitude is based on a biaxial ellipsoid shape with the Hapke scattering law and Kaasalainen law, as presented in Section \ref{sec:Scatter}. The relationship between amplitude and axis ratio is discussed in Section \ref{sec:Relation}, together with a comparison of phase effects in different asteroid spectral types. We briefly consider applications to data  in Section \ref{sec:Application}. \\

\section{Scattering Law} \label{sec:Scatter}

\subsection{Simulated Brightness}
To invert photometric observations \citep{2001Mkopti1, Muinonen2015AA},  the simulated brightness is calculated from the surface integral based on different shape models, scattering laws, and viewing geometries, using
\begin{equation}
\label{Eqn:Integral}
L =  \iint_{C+} S(\alpha, \mu, \mu_0) \, d \sigma.
\end{equation}
Here, $S(\alpha, \mu, \mu_0)$ is the scattering function and $C+$ represents the part of surface both illuminated and visible for different shape models, commonly considered as triaxial ellipsoids \citep{Cellino2015PSS}, convex polyhedron shapes \citep{LUXP2012AA}, and the intermediate Cellinoid shapes \citep{LUXP2016ICARUS, LUXP2017PSS}. 
$\mu, \mu_0$ in Equation (\ref{Eqn:Integral}) denote the projections of unit vectors of viewing and illuminating directions on the surface normal direction.

\subsection{Scattering Functions} 
\cite{Hapke2012Book} introduced a semi-physical model for illustrating the light reflection behaviour of a planetary surface. Here we exploit the five-parameter version for the scattering law, $S_{HM}(\alpha, \mu, \mu_0)$, written as
\begin{equation}
\label{Eqn:Hapke}
S_{HM}(\alpha, \mu, \mu_0)= 
\frac{\omega}{4}\frac{\mu_0}{\mu_0+\mu}\left\{\left[1+B(\alpha)\right]p(\alpha)+H(\mu_0)H(\mu)-1\right\}S(i,e,\alpha,\bar{\theta}),
\end{equation} 
where the five parameters are the single scattering albedo ($\omega$), roughness ($\bar{\theta}$), asymmetry factor ($g$), opposition surge amplitude ($B_{so}$), and opposition surge width ($h_s$), respectively. The related formulae, such as shadow hiding opposition surge function $B(\alpha)$, average particle single-scattering phase function $p(\alpha)$, and Ambartsumian-Chandrasekhar function for multiple scattering $H(x)$ can be found in   \cite{Hapke2012Book} (page 323).  \\

For simplicity, some empirical scattering functions are frequently used in numerical simulations. 
For example, \cite{2001MKopti2} used a three-parameter empirical phase function, which depends only on phase angle, with an empirical disk function (which depends solely on incidence and emission angles) to describe the surface scattering function, $S_{MK}(\alpha, \mu, \mu_0)$. The phase function is given by 
\begin{equation}
\label{Eqn:F}
f(\alpha) = A_0 \exp \left ( -\frac{\alpha}{D} \right ) + K \alpha + 1,
\end{equation}
where the opposition effect is described by a simple exponential term with height ($A_0$) and width ($D$) followed by a linear term with slope, $K$, to describe the phase curve behavior outside of opposition. The phase function is a combination of a single (Lommel-Seeliger function) and multiple (Lambert function) term, with a relative weighting factor, $C$, such that,
\begin{equation}
\label{Eqn:MKS}
S_{MK}(\alpha, \mu, \mu_0) = f(\alpha) \left(\frac{\mu \mu_0}{\mu + \mu_0} + C \mu\mu_0\right).
\end{equation}

\subsection{Asteroid Taxonomy}
Hapke photometric parameters representing asteroids of different spectral types are listed in Table \ref{Tab:CMP}. \cite{Li2015} listed the corresponding five parameters of the Hapke model for S-type, C-type, and V-type asteroids. The C-type asteroids have an average albedo $\omega = 0.037$, indicating a dark surface, while V-type asteroids have a brighter average albedo $\omega = 0.51$. The S-type asteroids are  intermediate, with the average albedo $\omega = 0.23$. \\

The Hapke model is difficult to invert numerically and so, instead, we fit the related parameters of the Kaasalainen model (MK Parameters) for different  asteroid spectral types. 
First, the simulated light curve ($LC_{HM}$) is generated based on one specified ellipsoid shape ($C$) with the supposed viewing($E$) and illuminating($E_0$) geometries, as well as the Hapke scattering parameters ($P_{HM}$) in Equation (\ref{Eqn:Hapke}) for some taxonomy as listed in Table \ref{Tab:CMP}. Each point in the light curve ($LC_{HM}$) corresponding to one rotational phase angle can be calculated by the surface integration using Equation (\ref{Eqn:Integral}) as follows,
\begin{equation}
LC^{(i)}_{HM} (C, E, E_0, P_{HM}) =   \iint_{C+} S_{HM}(\alpha, \mu, \mu_0) \, d \sigma.
\end{equation}

Then we search for the best-fit parameters of the Kaasalainen model by minimizing,
\begin{equation}
\chi^2 = \|| LC_{HM} - LC_{MK}\||_2^2
\end{equation}
where $LC_{MK}$ denotes the simulated light curve generated based on MK scattering parameters in Equation (\ref{Eqn:MKS}). Each point in $LC_{MK}$ is calculated under the same conditions as $LC_{HM}$, except for the scattering model,
\begin{equation}
LC^{(i)}_{MK} (C, E, E_0, P_{MK}) =   \iint_{C+} S_{MK}(\alpha, \mu, \mu_0) \, d \sigma.
\end{equation}
It should be noted that as the simulated lightcurves here are not relative (i.e. Normalization), the general form of the phase function is exploited as follows,
\begin{equation}
\label{Eqn:F2}
f(\alpha) = A_0 \exp \left ( -\frac{\alpha}{D} \right ) + K \alpha + B.
\end{equation}
with the additional back-ground intensity $B$ in Kaasalainen model, as presented in \cite{Kaasalainen2003}.

Finally, we test more than 1000 cases covering different shape models and observing geometries to obtain the distributions of MK parameters for each spectral type. The obtained MK parameters are listed in Table \ref{Tab:CMP}, together with the Hapke parameters for comparison.

Table \ref{Tab:CMP}  shows that as the albedo increases, the weight factor $C$ also increases, meaning that  multiple scattering effects  increase as the albedo increases (consistent with \cite{Hapke2012Book}). Furthermore, the parameter $C$ in the Kaasalainen scattering law is evidently related to the asteroid taxonomy. It can be applied to roughly classify the asteroids based on the derived MK parameters from photometric observation. In the numerical simulation, the appropriate $C$ value can improve the accuracy of other parameters, such as the pole direction and shape, if the asteroid types are known from  other measurements, such as infrared observations \citep{Wright2010AJ}.  \\

Furthermore,  multiple scattering is  intrinsic to the asteroid surface and, unlike the phase function, is not cancelled when calculating the amplitude. Therefore, for simplicity in the following section the relationship between amplitude and axis ratio for different taxonomical asteroids will adopt the derived $C$ in Table \ref{Tab:CMP}. \\

\section{Amplitude and Axis Ratio}\label{sec:Relation}

The observed brightness of an asteroid is related to the viewing and illumination geometries, as well as to  its physical parameters, including pole orientation and overall shape. Here we suppose that the shape is a biaxial ellipsoid ($a>b=c$) in rotation about the $c$-axis, and the viewing and illuminating directions are located on the equatorial plane of the asteroid. With this simplification, the simulated amplitude will provide an estimate of the upper limit to the amplitudes in different circumstances.\\

The lightcurve amplitude $\Delta m$ is calculated from

\begin{equation}
\Delta m = 2.5 \log \left (\frac{L_{\max}}{L_{\min}} \right )
\end{equation}

\noindent where $L_{\min}$  and  $L_{\max}$ are the minimum and maximum values of the calculated lightcurve, respectively.  

The Kaasalainen scattering law in Equation (\ref{Eqn:MKS}) is applied to calculate the simulated brightness in Equation (\ref{Eqn:Integral}). The primary contributions to the amplitude, $\Delta m$, are the disk function with the weight factor $C$ and the shape. \\


We computed the relationships between lightcurve amplitude, $\Delta m$, and  axis ratio, $(a/b)$,  with respect to phase angle for C-type,  S-type, and V-type asteroids.  The results for C-types are shown  in Figure \ref{Fig:C-type} while those for S-types are in Figure \ref{Fig:S-type}. The results for V-types are very similar and so are not separately plotted here.  Weidenschilling (1980) noted that a rotating fluid (strengthless) body cannot adopt an equilibrium shape if $a/b >$ 2.3.  Indeed, almost all well-observed asteroids have $a/b$ smaller than this value but, at small sizes where strength effects are important, larger ratios are possible.  With this  in mind, we consider a range of axis ratios from $a/b$ = 2:1 to $a/b$ = 7:1.   We extend  the models in Figures \ref{Fig:C-type} and \ref{Fig:S-type} only to $\alpha$ = 140\degr~because, at larger phase angles, the computed lightcurves become very sensitive to the adopted shape and scattering function and, in any case, essentially no observations exist at such large angles.

For small values of $a/b$, Figures \ref{Fig:C-type} and \ref{Fig:S-type} show that $\Delta m \propto \alpha$ for $\alpha \gtrsim 20$\degr, broadly  confirming the result of \cite{Zappala1990}. At   phase angles $\alpha \lesssim 20^\circ$, the amplitude is a slightly non-linear function of $\alpha$, as a result of the opposition effect. As the phase angle approaches $0^\circ$, the observed brightness will increase exponentially.  Comparison of Figures \ref{Fig:C-type} and \ref{Fig:S-type}  shows that the phase angle dependence of the amplitudes, for a given shape, is only weakly dependent on the asteroid spectral type. However, apparently the amplitude slightly increases as the albedo increases, especially as observed at the small phase angles.
\\

\section{Applications} \label{sec:Application}

An obvious conclusion from Figures \ref{Fig:C-type} and \ref{Fig:S-type} is that high phase angle observations will preferentially produce large amplitude lightcurves.  Measurements of the shapes and shape distributions of asteroids, if based on observations taken at non-zero phase angles, should be interpreted with appropriate caution. Likewise, rotational lightcurves accumulated over a wide range of phase angles can be compromised by the phase angle dependence, as is apparently the case in recent spacecraft observations of Kuiper belt objects 2012 HE$_{85}$ and 2011 HK$_{103}$ \citep{Verbiscer2019}. 

A less obvious conclusion is that the exaggerated lightcurves will lead to a size-dependent bias in asteroid surveys undertaken at large phase angles (e.g.~the small elongation/large phase angle  ``sweet spot'' survey strategy originally proposed for Panstarrs; Wainscoat et al.~2015).  Most asteroid sky surveys are magnitude-limited and objects whose mean magnitude is close to the survey detection limit will oscillate in and out of detection, according to the instantaneous rotational phase.  The result is a bias against the detection of faint asteroids, because they will be missing in many of the repeat visits needed by automatic detection routines for their identification.       For example, at phase angle $\alpha$ = 100\degr,  a body with unremarkable axis ratio $a/b$ = 2:1 would show a lightcurve amplitude $\Delta m \sim$ 1.8 magnitudes (a factor of 5).  Objects having a mean brightness within $\Delta m$ magnitudes of the survey limit would dip into and out of detection according to the rotational phase.  Bright asteroids (i.e.~those with mean magnitude brighter than the survey limit by an amount $>\Delta m$) can remain above the detection threshold regardless of rotational phase. Since brightness and asteroid size are (statistically) related, the under-counting of faint asteroids will lead to an apparent flattening of the size distribution, and an under-estimate of the flux of Earth-approaching bodies in asteroid surveys when conducted away from opposition. 

The models show that the angle dependence of the lightcurve amplitude is slightly influenced by multiple scattering and is therefore  related to the albedo and so to the spectral type. In principle, this fact could be used to  spectrally classify  asteroids even using single filter measurements taken over a range of geometries.  Such measurements might be produced, for example, by wide field sky surveys taken for other scientific purposes.  However, the spectral type dependence is very weak (compare Figures \ref{Fig:C-type} and \ref{Fig:S-type}) so that high quality observations would be needed for this single filter spectral-typing to be possible.  \\

\begin{figure}[ht!]
\fig{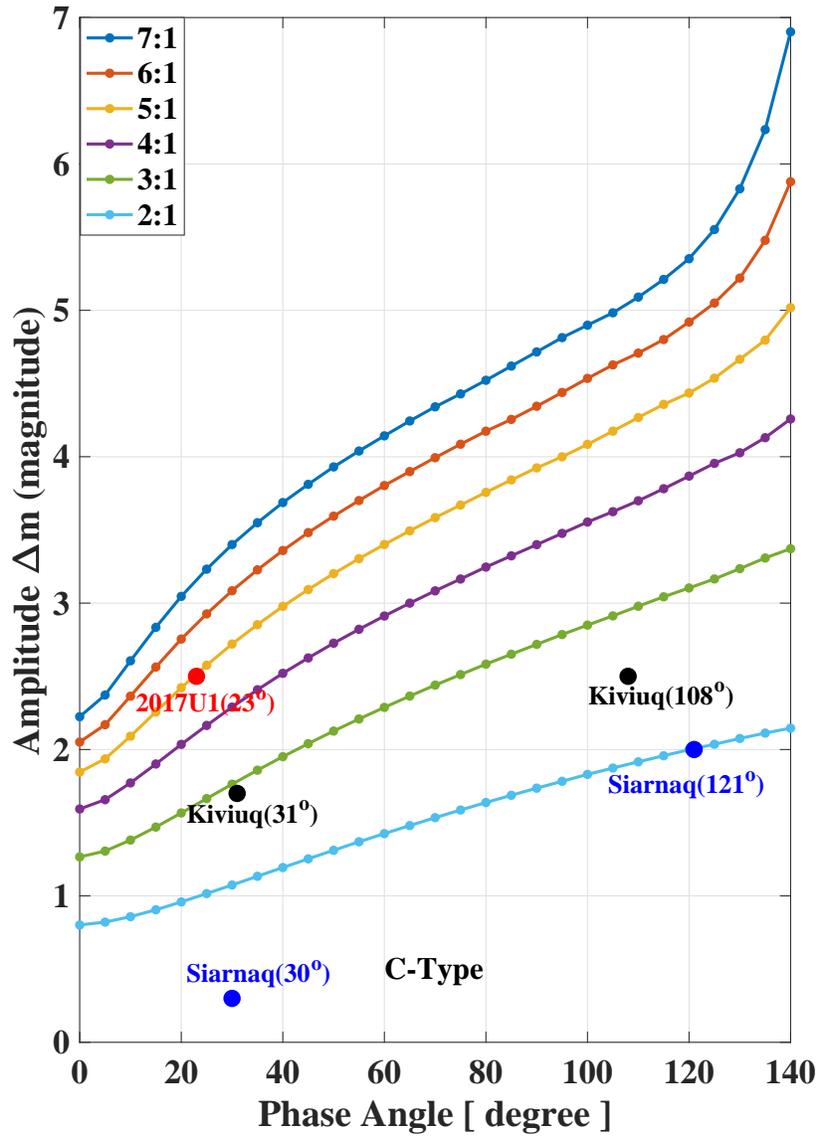}{0.8\textwidth}{}
\caption{Lightcurve amplitude, $\Delta m$ vs.~phase angle for C-type scattering, with the notations of 2017U1 and Saturn satellites, Kiviuq and Siarnaq, corresponding the observed phase angles. \label{Fig:C-typeU1}}
\end{figure}

To date, few high phase angle lightcurves appear in the literature.  
To give one example, \cite{Denk2018} presented  observations of the irregular satellites of Saturn, including  some high phase angle lightcurves. The two objects with the largest lightcurve amplitudes are plotted in Figure \ref{Fig:C-typeU1}.  Lightcurves of 15 km scale Kiviuq were determined at phase angle $\alpha = 108^\circ$ with amplitude  $\Delta m = 2.5$ magnitude, and 40 km scale Siarnaq was observed at   $\alpha = 121^\circ$ with $\Delta m = 2.0$  magnitude, respectively. These very large amplitudes, if interpreted geometrically,  would indicate axis ratios 10$^{0.4\Delta m}$ = 10:1 and 6:1, respectively.   Inspection of Figure \ref{Fig:C-typeU1} shows that inclusion of the phase angle dependence of the scattering gives a much more modest axis ratios from  $\sim$2.0 to 2.6.

As another example, the so-called interstellar asteroid, 1I/(2017 U1) `Oumuamua (hereafter ``U1''), displayed an abnormally large rotational amplitude, $\Delta m \sim 2.5$ magnitude that was interpreted as axis ratio $a/b \ge$ 10 by \cite{Meech2017nature}. However, U1 was not observed at $\alpha =0$\degr~and so the phase angle vs~amplitude effect must be taken into account.  Specifically, observations of U1 were possible over only a very narrow range of dates, with most  conducted  between October 25 and October 30, 2017 when the solar phase angle was $\alpha \sim 23^\circ$.  The albedo of this object could not be determined \citep{Trilling2018}, but a comet-like geometric albedo $p = 0.04$ is widely assumed \citep{Meech2017nature,Fitzsimmons2018Nature}.  With  $\alpha = 23^\circ$, $\Delta m$ = 2.5 magnitudes and $p$ = 0.04 (corresponding to a C-type surface) we can estimate  from  Figure \ref{Fig:C-type} an axis ratio $a/b =5$. This result is similar to the estimate by \cite{Bannister2017ApJL}, who gave an estimate for the axis ratio about $5.3:1$.  This is still a much larger axis ratio than found among the main-belt asteroids, most of which have $a/b <$ 2, but less extreme than obtained when neglecting the phase angle effect.\\

\begin{figure}[ht!]
\plotone{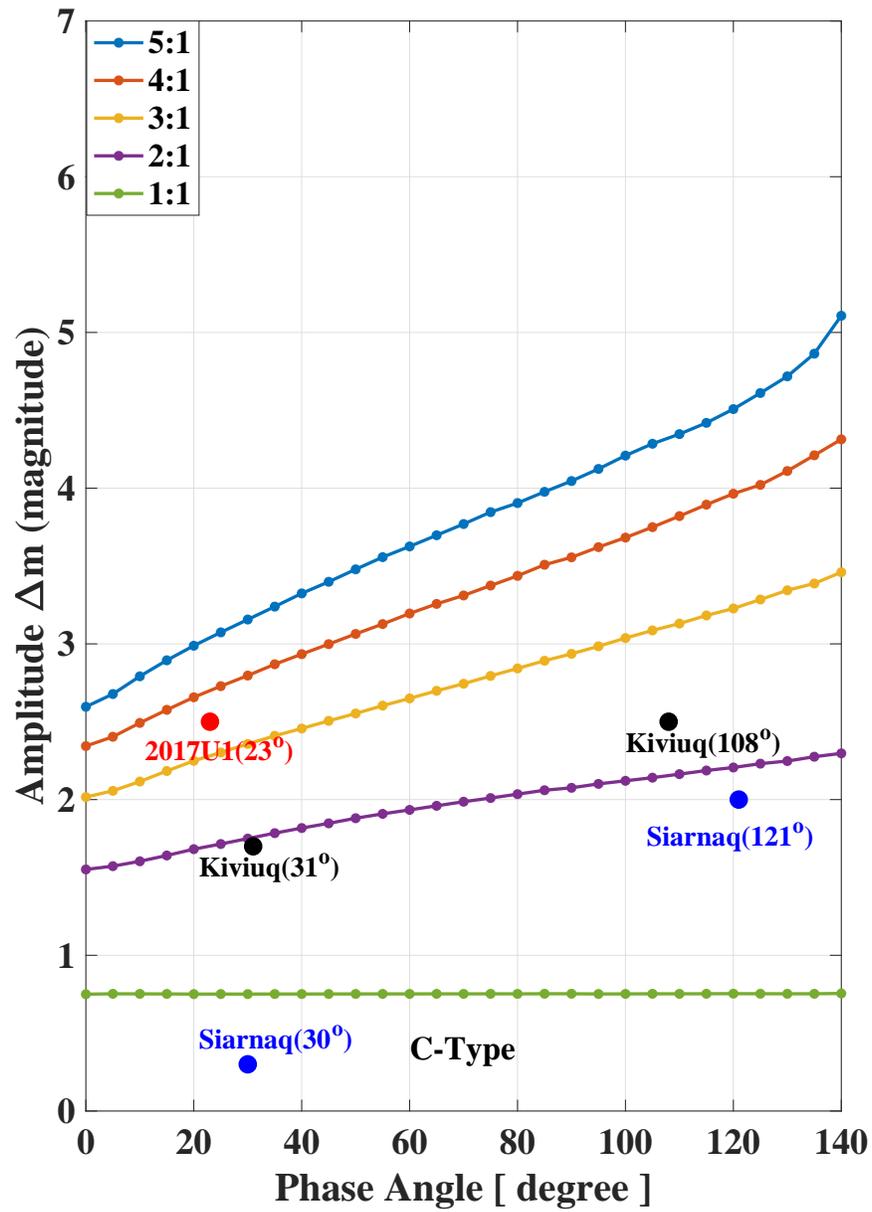}
\caption{Same as Figure \ref{Fig:C-typeU1} but for a contact binary model with  C-type scattering function. \label{Fig:Binary}}
\end{figure}

As an experiment, we also simulated the lightcurve for an assumed contact binary shape in which the two components are identical biaxial ellipsoids (c.f.~Lacerda and Jewitt (2007), who used Jacobi ellipsoids and different scattering functions).  The relationship between amplitude and  axis ratio is shown in Figure \ref{Fig:Binary}, where we have assumed a C-type surface with albedo $p = 0.037$. In this case, we find that the lightcurve of U1 can be approximated by nose-to-nose biaxial ellipsoids each with axis ratio $a/b = 3.5$, as marked in the figure. \\

\begin{figure}[ht!]
\gridline{\fig{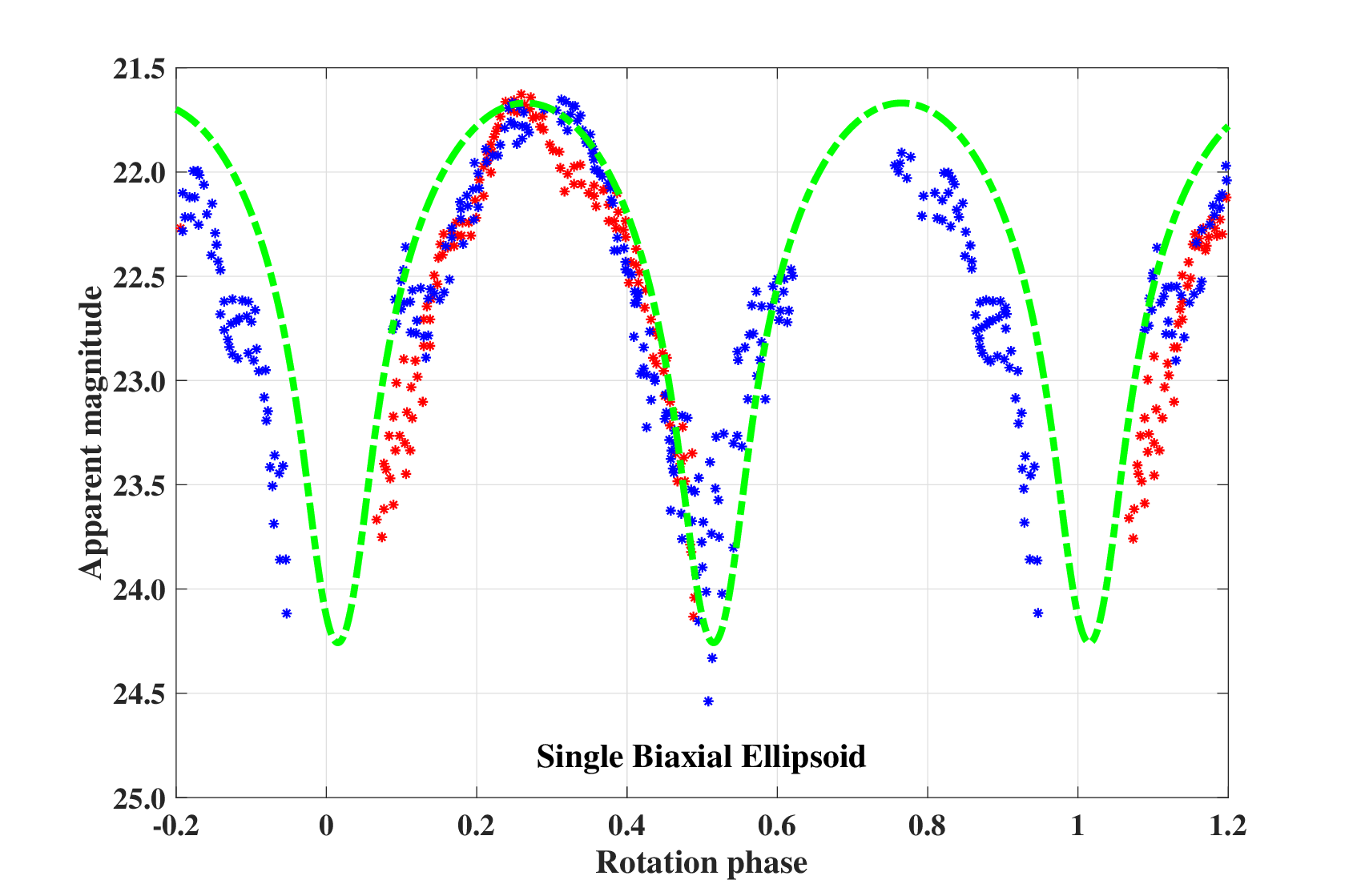}{0.7\textwidth}{}}
\gridline{\fig{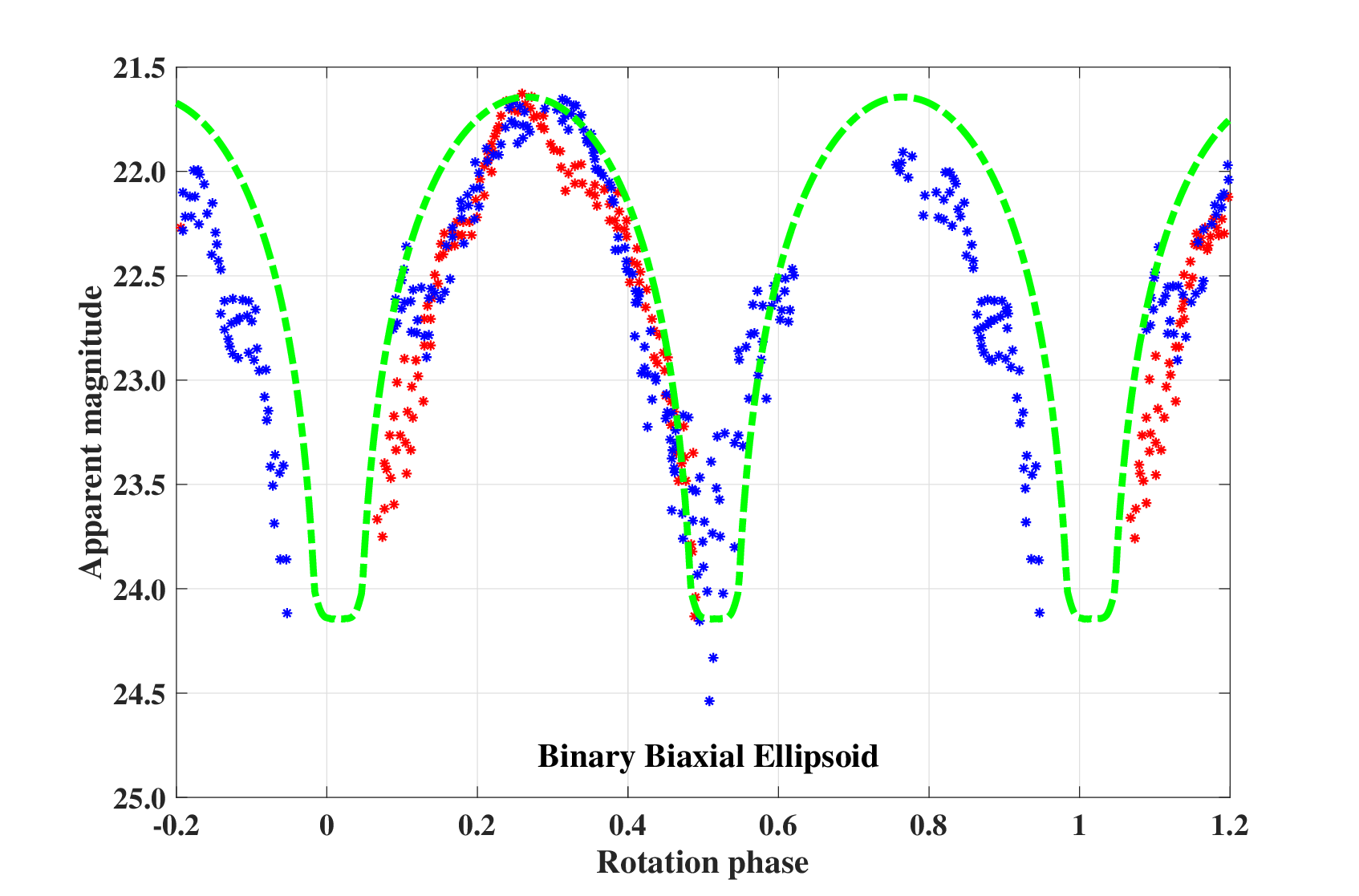}{0.7\textwidth}{}}
\caption{Simulated lightcurves (green line) for U1 at $\alpha$ = 23\degr~based on single and binary shape models, respectively, with the comparison to the observations (denoted by red and blue stars) by \cite{Drahus2018}. \label{Fig:LC}}
\end{figure}

Figure \ref{Fig:LC} shows  simulated lightcurves from the two different shape models, overplotting the lightcurve data from \cite{Drahus2018}. In the top panel we plot the single biaxial ellipsoid model with axis ratio $a/b = 5.2$ while  the lower panel shows a contact binary model  having equal components each with  $a/b = 3.5$. Both shape models can crudely reproduce the lightcurve amplitudes. We have made no attempt to match the detailed shape of the lightcurve because, in such a small body, the effects of irregularity are likely to be significant.  However, one feature of the lightcurve that is less sensitive to irregularity or other details of the shape is the flat minimum present in  the contact binaries caused by mutual shadowing (Figure \ref{Fig:LC} and Lacerda and Jewitt 2007).  The lightcurve of U1 does not show a flat minimum.  For this reason, we prefer the single ellipsoid model, and conclude that U1 has an axis ratio closer to $a/b \sim$ 5 than to $a/b \sim$ 10, in agreement with Drahus et al.~(2018).\\

\clearpage
\section{Summary} \label{sec:Conclusion}

\begin{figure}[ht!]
\fig{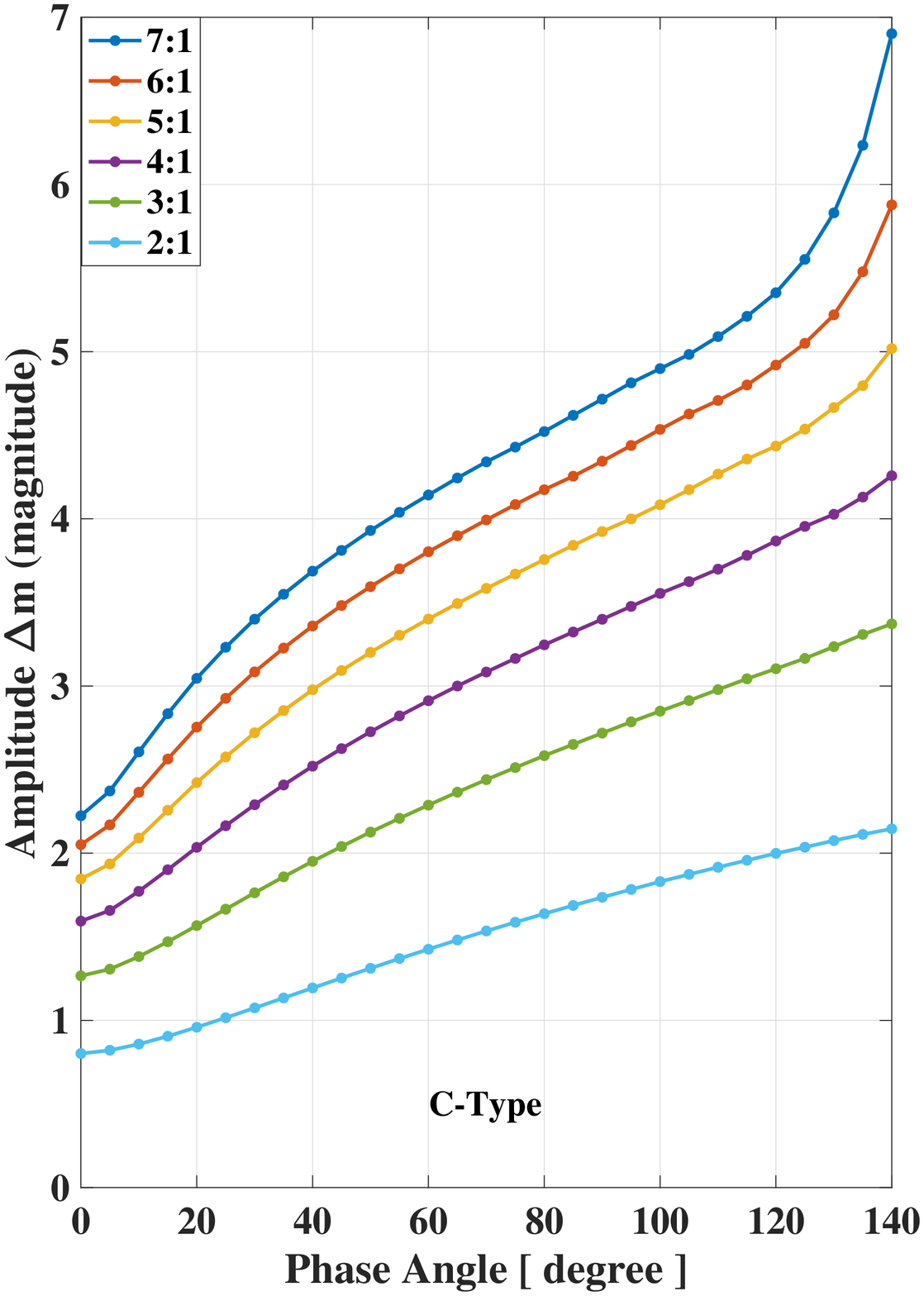}{0.8\textwidth}{}
\caption{Lightcurve amplitude, $\Delta m$ vs.~phase angle  $0^\circ \le \alpha \le 140^\circ$ for biaxial ellipsoid ($a>b=c$) models with  $a/b$ =2,3...7,   and with  C-type scattering, assumed average albedo $p = 0.037$. \label{Fig:C-type}}
\end{figure}

\begin{figure}[ht!]
\fig{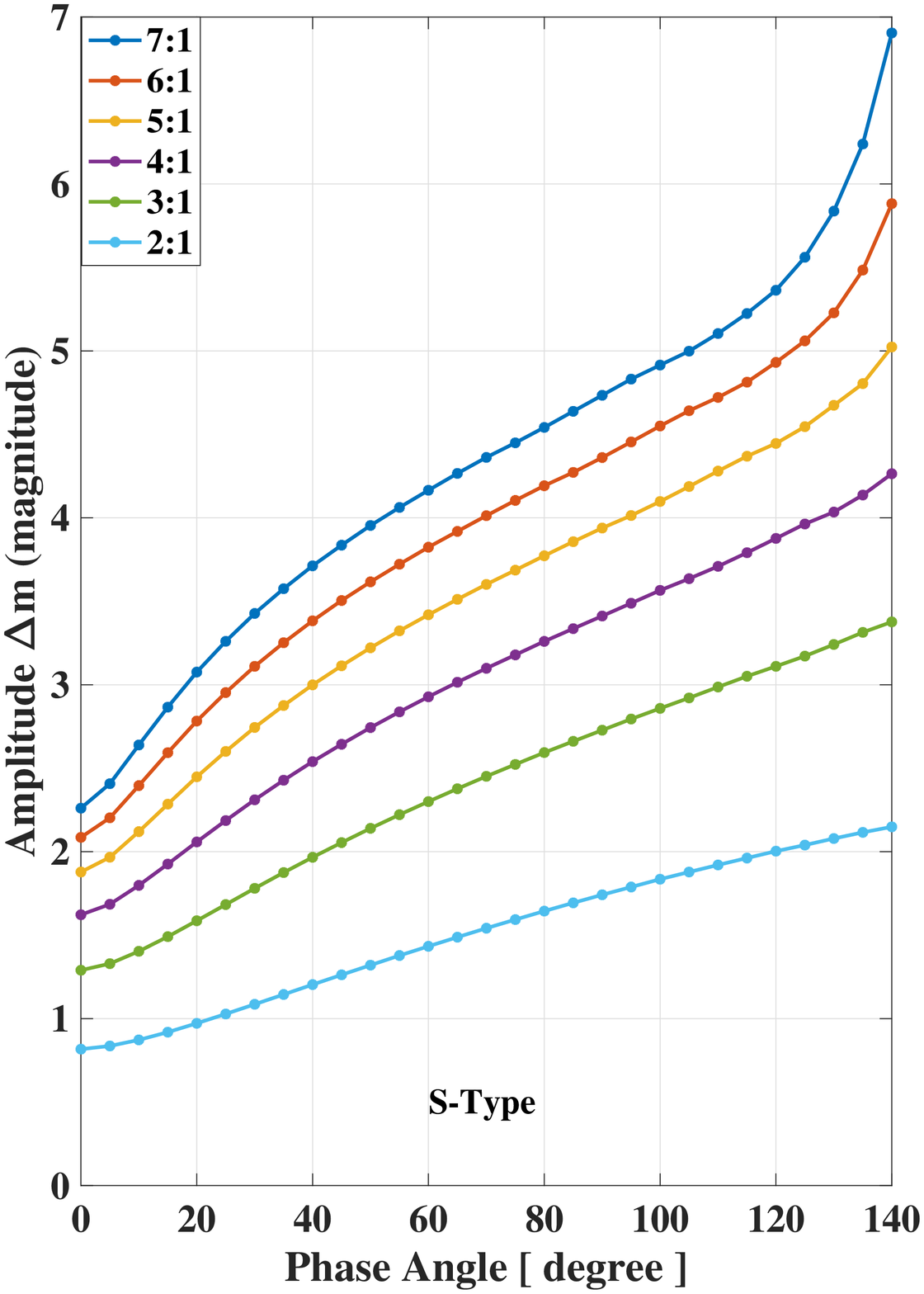}{0.8\textwidth}{}
\caption{Lightcurve amplitude, $\Delta m$ vs.~phase angle  $0^\circ \le \alpha \le 140^\circ$ for  biaxial ellipsoid ($a>b=c$) models with  $a/b$ =2,3...7,   and with  S-type scattering, assumed average albedo $p = 0.23$. \label{Fig:S-type}}
\end{figure}

\begin{itemize}
\item We present models of the phase angle dependence of the lightcurve amplitude for asteroids having different axis ratios, spectral types, and ellipsoid and contact binary body-shapes. Our results (Figures \ref{Fig:C-type} and \ref{Fig:S-type}) provide a quick constraint on the shape for observations taken at non-zero phase angles. 

\item We find that, to first order, the amplitude vs.~phase relation is independent of the assumed asteroid albedo and spectral type, and depends only on the body shape.  


\item Using the  observations of 1I/2017 U1 `Oumuamua tabulated by Drahus et al.~(2018), we infer axis ratio $5.2 : 1$ for an ellipsoid shape model.  We can also match the amplitude with a contact binary model in which each component has axis ratio $3.5 : 1$.  However, the lightcurve of U1 lacks the flat-bottomed minima characteristic of contact binaries and so we prefer the former representation.

\end{itemize}

\acknowledgments
We thank the anonymous referees for their constructive comments. XL is funded by The Science and Technology Development Fund, Macau SAR (File no. 0018/2018/A).
\bibliographystyle{aasjournal}


\clearpage

\begin{deluxetable*}{ccccc}
\tablenum{1}
\tablecaption{Derived MK Parameters from Hapke Model for different asteroids\label{Tab:CMP}}
\tablewidth{0pt}
\tablehead{
\colhead{Parameters} & \colhead{C-Type} & \colhead{S-Type} & \colhead{V-Type}
}
\startdata
\multicolumn{4}{c}{Hapke Model}\\
          \noalign{\smallskip}\hline
$\omega$ & 0.037 & 0.23 & 0.51 \\
$g$ & 1.03 & 1.6 & 1 \\
$B_{so}$ & 0.025 & 0.08 & 0.098 \\
$h_s$ & -0.47 & -0.27 & -0.26\\
$\bar{\theta}$ & 20 & 20 & 32 \\
          \noalign{\smallskip}\hline\hline
 \multicolumn{4}{c}{Kaasalainen Model}\\
          \noalign{\smallskip}\hline
$A_0$ & $0.009 \sim 0.01$ & $0.039 \sim 0.041$ & $0.046 \sim 0.048$ \\
$D$ & $3.3 \sim 3.5$ & $8.4 \sim 8.6$ & $9.2 \sim 9.45$  \\
$K$ & $-2.4 \sim -2.2$\, $ \times 10^{-4}$ & $-4.8 \sim -4.6$ \, $\times 10^{-4}$ & $-7.7 \sim -7.3$ \, $\times 10^{-4}$  \\
$B$ & $0.0142 \sim 0.0148$ & $0.049 \sim 0.051$ & $0.081 \sim 0.087$ \\
$C$ & $0.08 \sim 0.11$& $0.11 \sim 0.15$ & $0.28 \sim 0.34$ \\
          \noalign{\smallskip}\hline
\enddata
\end{deluxetable*}

\clearpage

\clearpage

\clearpage

\clearpage

\clearpage

\end{document}